\documentclass[a4paper, 11pt]{article}
\usepackage{jheppub}\makeatletter\gdef\@fpheader{}\makeatother
\usepackage{mathtools, bbm, microtype, tabularx, subfigure, graphicx, placeins}
\usepackage[dvipsnames]{xcolor}
\usepackage{youngtab}\Yboxdim 6.5pt\Ylinethick 0.4pt

\graphicspath{{Figures/}}

\newcommand{\Abs}[1]{\left\lvert#1\right\rvert}
\newcommand{\RealPart}[1]{\textrm{Re}(#1)}
\newcommand{\ImaginaryPart}[1]{\textrm{Im}(#1)}
\newcommand{\U}{\textrm{U}}
\newcommand{\SU}{\textrm{SU}}
\newcommand{\USp}{\textrm{USp}}  
\def\T{\mathbbm{T}}
\def\Z{\mathbbm{Z}}

\author[a,b,c]{Adeel Mansha,}
\author[d,e]{Tianjun Li,}
\author[f]{Mudassar Sabir\footnote{Corresponding author}} 

\affiliation[a]{Department of Physics, Zhejiang Normal University, Jinhua 321004, P. R. China}
\affiliation[b]{Zhejiang Institute of Photoelectronics, Zhejiang Normal University, Jinhua 321004, P. R. China}
\affiliation[c]{Centre For High Energy Physics, University of the Punjab, Lahore 54590, Pakistan}
\affiliation[d]{CAS Key Laboratory of Theoretical Physics, Institute of Theoretical Physics, Chinese Academy of Sciences, Beijing 100190, P. R. China}
\affiliation[e]{School of Physical Sciences, University of Chinese Academy of Sciences, Beijing, P. R. China}
\affiliation[f]{School of Physics, University of Electronic Science and Technology of China, Chengdu, Sichuan 611731, P. R. China} 
	
\emailAdd{adeelmansha@zjnu.edu.cn}
\emailAdd{tli@itp.ac.cn}
\emailAdd{mudassar.sabir@uestc.edu.cn} 

\arxivnumber{}

\begin{document}

\title{Revisiting the ${\cal N}=1$ supersymmetric trinification models from intersecting D6-branes}

\abstract{We revisit the construction of the $\mathcal{N}=1$ supersymmetric trinification models with gauge group $\mathrm{U}(3)_C\times \mathrm{U}(3)_L \times \mathrm{U}(3)_R$ in Type IIA string theory on $\mathbf{T^6/(\mathbb{Z}_2\times \mathbb{Z}_2)}$ orientifold with intersecting D6-branes. The non-trivial K-theory conditions and tadpole cancellation conditions severely restrict the number of allowed models even for the case of rectangular two-tori. Using a supervised search algorithm, we find a few four-family models where the highest wrapping number is 2. For these models, we present the complete particle spectra and the gauge coupling relations at the string-scale. }

\maketitle
\section{Introduction}\label{sec:Intro}

The idea of unifying the Standard Model (SM) gauge interactions into a Grand Unified Theory (GUT) has an old history \cite{Salam:1980jd}. In order for such a GUT to be both tenable and internally consistent, it necessitates a symmetry group characterized by a simple Lie group, one that encompasses the SM's group $\SU(3)_C\times \SU(2)_L \times \U(1)_Y$ as a subgroup, and crucially, permits complex representations to enable the emergence of a chiral theory. The fundamental examples within the unitary and orthogonal group families are the well-recognized $\SU(5)$ and $\textrm{SO}(10)$ groups. Later, with the advent of heterotic string theory, the gauge group $\textrm{E}_{6} \subset \textrm{E}_{8}$ has become more important. Interestingly, the trinification group $\SU(3) \times \SU(3) \times \SU(3)$ is also one of the subgroups of $\textrm{E}_{6}$ which has recently gained renewed interest from $\textrm{E}_6$ symmetry breaking. In the trinification models, the quarks, leptons and Higgs fields all arise from the bifundamental representations of the gauge group $\SU(3)_C\times \SU(3)_L \times \textrm{SU}(3)_R$. 

Apart from choosing the GUT gauge group, embedding the SM into string theory also involves a certain choice of Calabi-Yau compactification. The SM fermions belong to the chiral representations of the gauge group $\SU(3)_C\times \SU(2)_L \times \U(1)_Y$ such that all the gauge anomalies are canceled. In Type II string theory, simply placing the parallel D-branes in flat space does not yield chiral fermions. Instead, to realize the chiral fermions we need to consider either the D-branes on orbifold singularities \cite{Aldazabal:2000sa} or the intersecting D-branes on generalized orbifolds called orientifolds \cite{Shiu:1998pa, Cvetic:2001tj}. In orientifolds, both the discrete internal symmetries of the world-sheet theory and the products of internal symmetries with world-sheet parity reversal become gauged. In Type IIA string theory, setting the intersecting D6-branes over orientifolds provides a geometric framework to generate the gauge structures, chiral spectra, and various couplings. In particular, the four-dimensional gauge couplings depend on the volume of the cycles wrapped by the D6-branes, and the gravitational coupling is determined by their total internal volume. Similarly, the cubic Yukawa couplings depend exponentially on the triangular areas of open worldsheet intersections which naturally explain the observed fermion mass hierarchies. The general flavor structure and selection rules for intersecting D6-brane models have been investigated in \cite{Chamoun:2003pf, Higaki:2005ie}. 

In Ref.~\cite{Chen:2005mj} one of us (TL) had constructed the first consistent three-family supersymmetric trinification model with gauge group $\SU(3)_C \times \SU(3)_L \times \SU(3)_R$ from intersecting D6-branes over Type IIA $\T^6/(\Z_2\times \Z_2)$ orientifold. It was pointed out that unlike the Pati-Salam gauge group, the RR tadpole cancellation and K-theory conditions for the trinification gauge group are much more difficult to satisfy. More recently, in the context of supersymmetric Pati-Salam models, the complete landscape with three families \cite{He:2021gug, Mansha:2023kwq, Sabir:2022hko, He:2021kbj, Mansha:2022pnd} and four families \cite{Li:2022fzt} has already been enumerated. In fact, the SM does not provide a theoretical prediction regarding the number of families. Allowing for a fourth family naturally leads to the flavor democracy hypothesis as discussed in~\cite{Sultansoy:2006dg}. The particles from the massive fourth generation would alter the cross sections and the decay channels of the Higgs particles and when the Yukawa couplings of the fourth generation particles are large enough, these particles are natural candidates for electroweak symmetry breaking \cite{Hung:2010xh}. 

In this paper we present results of the four-family supersymmetric trinification models from the intersecting D6-branes on a Type IIA $\T^6/(\Z_2\times \Z_2)$ orientifold. 
The plan of the paper is as follows. In section~\ref{sec:orientifold}, we review the model building rules for $\mathcal{N}=1$ supersymmetric trinification models with  gauge group $\SU(3)_C \times \SU(3)_L \times \SU(3)_R$ on a $\T^6/(\Z_2\times \Z_2)$ orientifold. In section~\ref{sec:brane-splitting} we discuss the gauge symmetry breaking via the brane-splitting mechanism to yield the SM gauge group. In section~\ref{sec:Pheno} we discuss the salient phenomenological features and comment on the decoupling of the various exotic particles from the low energy spectra through strong dynamics in a particular model. Finally, we conclude in section~\ref{sec:conclusion}.

\section{Trinification model building from $\T^6/(\Z_2\times \Z_2)$ orientifold} \label{sec:orientifold}

In the orientifold $\T^6/(\Z_2\times \Z_2)$, $\T^6$ is treated as a product of three two-tori with the orbifold group $(\Z_2\times \Z_2)$, having the generators $\theta$ and $\omega$ that are associated with the twist vectors $(1/2,-1/2,0)$ and $(0,1/2,-1/2)$ respectively such that their action on $z_i$ is given by,
\begin{align}
\theta: \quad (z_1,z_2,z_3) &\to (-z_1,-z_2,z_3), \nonumber \\
\omega: \quad (z_1,z_2,z_3) &\to (z_1,-z_2,-z_3). \label{orbifold}
\end{align}
Orientifold projection is the gauged $\Omega {\cal R}$ symmetry, where $\Omega$ denotes the world-sheet parity that interchanges the left- and right-movers of a closed string and acts on the ends of an open string as,
\begin{align}
\textrm{Closed}:  &\quad \Omega : (\sigma_1, \sigma_2) \mapsto (2\pi -\sigma_1, \sigma_2), \nonumber \\
\textrm{Open}:  &\quad  \Omega : (\tau, \sigma) \mapsto (\tau, \pi - \sigma) ,
\end{align}
and ${\cal R}$ represents the complex conjugation which acts on the complex coordinates $z_i$. This leads to the generations of four different kinds of orientifold 6-planes (O6-planes) related to $\Omega {\cal R}$, $\Omega {\cal R}\theta$, $\Omega {\cal R}\omega$, and $\Omega {\cal R}\theta\omega$ respectively. These orientifold projections are only consistent with either the rectangular or the tilted complex structures of the factorized two-tori. The wrapping numbers for the rectangular and tilted two-tori are denoted as $n_a^i[a_i]+m_a^i[b_i]$ and $n_a^i[a'_i]+m_a^i[b_i]$ respectively, where $[a_i']=[a_i]+\frac{1}{2}[b_i]$. Then for the rectangular and tilted two-torus, generic 1-cycle satisfies $l_{a}^{i}\equiv m_{a}^{i}$, and $l_{a}^{i}\equiv 2\tilde{m}_{a}^{i}=2m_{a}^{i}+n_{a}^{i}$ respectively. The homology cycles for a stack $a$ of $N_a$ D6-branes along the cycle $(n_a^i,l_a^i)$ and their $\Omega {\cal R}$ images ${a'}$ are defined as,
\begin{align}
[\Pi_a ]&=\prod_{i=1}^{3}\left(n_{a}^{i}[a_i]+2^{-\beta_i}l_{a}^{i}[b_i]\right), \nonumber \\
[\Pi_{a'}] &=\prod_{i=1}^{3}\left(n_{a}^{i}[a_i]-2^{-\beta_i}l_{a}^{i}[b_i]\right),
\end{align}
where $\beta_i=0$ or $\beta_i=1$ for the rectangular or tilted $i^\textrm{th}$ two-tori respectively. 

The homology 3-cycles wrapped by the four O6-planes are defined as
\begin{alignat}{2}
&\Omega {\cal R} :            &         [\Pi_{\Omega {\cal R}}] &= 2^3 [a_1]\times[a_2]\times[a_3],  \nonumber\\
&\Omega {\cal R}\omega :      &        [\Pi_{\Omega {\cal R}\omega}] &=-2^{3-\beta_2-\beta_3}[a_1]\times[b_2]\times[b_3],  \nonumber\\
&\Omega {\cal R}\theta\omega : \quad & [\Pi_{\Omega {\cal R}\theta\omega}] &=-2^{3-\beta_1-\beta_3}[b_1]\times[a_2]\times[b_3], \nonumber\\
&\Omega {\cal R}\theta :      &       [\Pi_{\Omega {\cal R} \theta}] &=-2^{3-\beta_1-\beta_2}[b_1]\times[b_2]\times[a_3]. \label{orienticycles}
\end{alignat}
The intersection numbers can be calculated in terms of wrapping numbers as,
\begin{align}
I_{ab} &=[\Pi_a][\Pi_b] =2^{-k}\prod_{i=1}^3(n_a^il_b^i-n_b^il_a^i),\nonumber\\
I_{ab'}&=[\Pi_a]\left[\Pi_{b'}\right] =-2^{-k}\prod_{i=1}^3(n_{a}^il_b^i+n_b^il_a^i),\nonumber\\
I_{aa'}&=[\Pi_a]\left[\Pi_{a'}\right] =-2^{3-k}\prod_{i=1}^3(n_a^il_a^i),\nonumber\\
I_{aO6}&=[\Pi_a][\Pi_\textrm{O6}] =2^{3-k}(-l_a^1l_a^2l_a^3+l_a^1n_a^2n_a^3+n_a^1l_a^2n_a^3+n_a^1n_a^2l_a^3),\label{intersections}
\end{align}
where $k=\sum_{i=1}^3\beta_i$ and $[\Pi_\textrm{O6}]=[\Pi_{\Omega {\cal R}}]+[\Pi_{\Omega {\cal R}\omega}]+[\Pi_{\Omega {\cal R}\theta\omega}]+[\Pi_{\Omega {\cal R}\theta}]$.

\subsection{Constraints from tadpole cancellation and supersymmetry}\label{subsec:constraints}

As D6-branes and O6-orientifold planes contain the Ramond-Ramond charges, therefore they must satisfy the Gauss's law in compact space implying the sum of D-brane and cross-cap RR-charges must vanish \cite{Gimon:1996rq}
\begin{eqnarray}\label{RRtadpole}
\sum_a N_a [\Pi_a]+\sum_a N_a \left[\Pi_{a'}\right]-4[\Pi_\textrm{O6}]=0,
\end{eqnarray}
where the last contribution comes from the O6-planes, which have $-4$ RR charges in D6-brane charge units. The constraint from RR tadpole is sufficient to cancel the $\SU(N_a)^3$ cubic non-Abelian anomaly, while Green-Schwarz mechanism, mediated by untwisted RR fields \cite{Green:1984sg}, plays a crucial rule in canceling the U(1) mixed gauge and gravitational or $[\textrm{SU}(N_a)]^2 \U(1)$ gauge anomaly.

For the sake of convenience, let us define the following products of wrapping numbers,
\begin{alignat}{4}
A_a &\equiv -n_a^1n_a^2n_a^3, &\quad B_a &\equiv n_a^1l_a^2l_a^3, &\quad C_a &\equiv l_a^1n_a^2l_a^3, \quad & D_a &\equiv l_a^1l_a^2n_a^3, \nonumber\\
\tilde{A}_a &\equiv -l_a^1l_a^2l_a^3, & \tilde{B}_a &\equiv l_a^1n_a^2n_a^3, & \tilde{C}_a &\equiv n_a^1l_a^2n_a^3, & \tilde{D}_a &\equiv n_a^1n_a^2l_a^3.\,\label{variables}
\end{alignat}
Cancellation of RR tadpoles requires the orientifold plane numbers also called ``filler branes'' that satisfy, trivially, the four-dimensional ${\cal N}=1$ supersymmetry conditions. Thus, the no-tadpole condition is given as,
\begin{align}\label{notadpoles}
 -2^k N^{(1)}+\sum_a N_a A_a &= -2^k N^{(2)}+\sum_a N_a B_a = \nonumber\\
 -2^k N^{(3)}+\sum_a N_a C_a &= -2^k N^{(4)}+\sum_a N_a D_a = -16,\,
\end{align}
where $2 N^{(i)}$ represents the number of filler branes wrapping along the $i^\textrm{th}$ O6-plane. The filler branes belong to the hidden sector $\USp$ group and carry the same wrapping numbers as one of the O6-planes as shown in table~\ref{orientifold}. $\USp$ group is hence referred with respect to the non-zero $A$, $B$, $C$ or $D$-type.

\begin{table}[th]
\renewcommand{\arraystretch}{1.3}
\caption{The wrapping numbers for four O6-planes.} \label{orientifold}
\begin{center}
$\begin{array}{|c|c|c|}
\hline
\text{Orientifold action} & \text{O6-plane} & (n^1,l^1)\times (n^2,l^2)\times (n^3,l^3)\\
\hline\hline
    \Omega {\cal R}& 1 & (2^{\beta_1},0)\times (2^{\beta_2},0)\times (2^{\beta_3},0) \\
\hline
    \Omega {\cal R}\omega& 2& (2^{\beta_1},0)\times (0,-2^{\beta_2})\times (0,2^{\beta_3}) \\
\hline
    \Omega {\cal R}\theta\omega& 3 & (0,-2^{\beta_1})\times (2^{\beta_2},0)\times (0,2^{\beta_3}) \\
\hline
    \Omega {\cal R}\theta& 4 & (0,-2^{\beta_1})\times (0,2^{\beta_2})\times (2^{\beta_3},0) \\
\hline
\end{array}$
\end{center}
\end{table}
Preserving ${\cal N}=1$ supersymmetry in four dimensions shows that the rotation angle of any D6-brane with respect to the orientifold plane must be an element of $\SU(3)$, i.e.
\begin{equation}
\theta^a_1 + \theta^a_2 + \theta^a_3 = 0 \mod 2\pi ,
\end{equation}
with $\theta^a_j = \arctan (2^{- \beta_j} \chi_j l^a_j/n^a_j)$. $\theta_i$ is the angle between the $D6$-brane and orientifold-plane in the $i^\textrm{th}$ two-tori and $\chi_i=R^2_i/R^1_i$ are the complex structure moduli for the $i^\textrm{th}$ two-tori.
${\cal N}=1$ supersymmetry conditions are given as,
\begin{eqnarray}
x_A\tilde{A}_a+x_B\tilde{B}_a+x_C\tilde{C}_a+x_D\tilde{D}_a=0,\nonumber\\
\frac{A_a}{x_A}+\frac{B_a}{x_B}+\frac{C_a}{x_C}+\frac{D_a}{x_D} < 0, \label{susyconditions}
\end{eqnarray}
where $x_A=\lambda,\; x_B=2^{\beta_2+\beta_3}\cdot\lambda /\chi_2\chi_3,\; x_C=2^{\beta_1+\beta_3}\cdot\lambda /\chi_1\chi_3,\; x_D=2^{\beta_1+\beta_2}\cdot\lambda /\chi_1\chi_2$.

Orientifolds also contain the discrete D-brane RR charges represented by the $\Z_2$ K-theory groups, which are invisible by the ordinary homology~\cite{Witten:1998cd, Cascales:2003zp, Marchesano:2004yq, Marchesano:2004xz}, should also be taken into account~\cite{Uranga:2000xp} during model building. The K-theory conditions read~\cite{Chen:2005mj},
\begin{eqnarray}
\sum_a N_a {A}_a  = \sum_a  N_a  B_a = \sum_a  N_a  C_a = \sum_a  N_a D_a = 0 \mod 4 \label{K-charges}~.
\end{eqnarray}

\begin{table}[th]
\renewcommand{\arraystretch}{1.3}\centering
\caption{General spectrum for intersecting D6-branes at generic angles, where ${\cal M}$ is the multiplicity, and  $a_{\protect\yng(2)}$ and $a_{\protect\yng(1,1)}$ denote respectively the symmetric and antisymmetric representations of $\U(N_a/2)$. Positive intersection numbers in our convention refer to the left-handed chiral supermultiplets.\\}
$\begin{array}{|c|c|}
\hline
\text{\bf Sector} & \phantom{more space inside this box}{\bf Representation} \phantom{more space inside this box} \\
\hline\hline
aa                & \U(N_a/2) \text{ vector multiplet}  \\
                  & \text{3 adjoint chiral multiplets}  \\
\hline ab+ba      & {\cal M}(\frac{N_a}{2}, \frac{\overline{N_b}}{2})= I_{ab}(\yng(1)_{a},\overline{\yng(1)}_{b}) \\
\hline ab'+b'a    & {\cal M}(\frac{N_a}{2}, \frac{N_b}{2})=I_{ab'}(\yng(1)_{a},\yng(1)_{b}) \\
\hline aa'+a'a    & {\cal M} (a_{\yng(2)})= \frac{1}{2} (I_{aa'} - \frac{1}{2} I_{aO6}) \\
                  & {\cal M} (a_{\yng(1,1)_{}})= \frac{1}{2} (I_{aa'} + \frac{1}{2} I_{aO6})  \\
\hline
\end{array}$
\label{tab:spectrum}
\end{table}
The multiplicities of the particles for intersecting D6-branes models are shown in table \ref{tab:spectrum}. By following the convention of \cite{Cvetic:2004ui}, the $N$ number of D6-brane stacks corresponds to U($N/2$) and USp($N$) respectively. A positive intersection number in our convention refers to the left-chiral supermultiplet, and vice versa.

Within trinification models, quarks, leptons, and the Higgs boson all originate from the bifundamental representations of the gauge group $\SU(3)_C\times \SU(3)_L \times \textrm{SU}(3)_R$. 
Accordingly, we require three stacks of six D6-branes to yield three $\SU(3)$ gauge group factors.
If the number of generations is $n_g$, and $n_h$ be the number of higgs, then we have,
\begin{align} 
n_g &= I_{ab} + I_{ab'} = - (I_{ac}+I_{ac'}), \nonumber \\
n_h &= \Abs{I_{bc}+I_{bc'}} - n_g . \label{eq:HiggsNumber}
\end{align}

\subsection{Gauge symmetry breaking}\label{sec:brane-splitting}

The $\U(3)_C\times \U(3)_L\times \U(3)_R$ gauge symmetry is broken down to the $\SU(3)_C\times \SU(3)_L\times \SU(3)_R$ gauge symmetry via the Green-Schwartz mechanism \cite{Aldazabal:2000dg, Ibanez:2001nd, Cvetic:2001nr}. The resulting $\SU(3)_C\times \SU(3)_L\times \SU(3)_R$ gauge symmetry can be broken down to the $\SU(3)_C \times \SU(2)_L \times \U(1)_{Y_L} \times \U(1)_{I_{3R}} \times \U(1)_{Y_R}$ gauge symmetry by the splittings of the $U(3)_L$ and $U(3)_R$ stacks of the D6-branes.

The SM electromagnetic charge generator $Q_{Y}$ is defined as
\begin{equation}\label{eq:QY}
Q_{Y} \equiv I_{3L} - \frac{Y_L}{2} + I_{3R} - \frac{Y_R}{2} ,
\end{equation}
where the generators for $\U(1)_{I_{3L}}$ and $\U(1)_{I_{3R}}$, and $\U(1)_{Y_L}$ and $\U(1)_{Y_R}$ in $\SU(3)_L$ and $\SU(3)_R$ gauge symmetries are
\begin{equation}
\mathbf{T}_{\U(1)_{I_{3L,R}}}= \left(  \begin{array}{ccc}
\frac{1}{2} & 0 & 0 \\
0 & -\frac{1}{2} & 0 \\
0 & 0 & 0  \end{array} \right)~,~\,
\end{equation}
\begin{equation}
\mathbf{T}_{\U(1)_{Y_{L,R}}}= \left(  \begin{array}{ccc}
\frac{1}{3} & 0 & 0 \\
0 & \frac{1}{3} & 0 \\
0 & 0 & -\frac{2}{3}  \end{array} \right)~.~\,
\end{equation}

And the explicit particle components in the bifundamental quarks $Q^i_{L,R}$, leptons $L^i$, and higgs $H^i$ representations are
\begin{align}
(\mathbf{3}, \mathbf{\bar{3}}, \mathbf{1}) :             \;\;
Q^i_L &= \left(\begin{array}{ccc}
d & u & h \\
d & u & h \\
d & u & h  \end{array} \right) ~,~\,\\
(\mathbf{\bar{3}}, \mathbf{1}, \mathbf{3}) :             \;\;
Q^i_R &= \left(\begin{array}{ccc}
d^c & d^c & d^c \\
u^c & u^c & u^c \\
h^c & h^c & h^c  \end{array} \right) ~,~\,\\
(\mathbf{1}, \mathbf{3}, \mathbf{\bar{3}}) :             \;\; 
L^i \; \mathrm{or} \; H^k &= \left(\begin{array}{ccc}
N & E^c & \nu \\
E & N^c & e \\
\nu^c & e^c & S  \end{array} \right) ~.~\,
\end{align}

Giving VEVs to the singlet Higgs fields $\nu^c$ and  $S$, we can break the $\U(1)_{Y_L}\times \U(1)_{I_{3R}}\times \U(1)_{Y_R}$ gauge symmetry down to the $\U(1)_Y$ hypercharge interaction.
\begin{equation}
\langle \nu^c \rangle \neq 0 , \qquad \langle S \rangle \neq 0 .
\end{equation}
The electric charges for $h$ and $h^c$ are respectively $-\frac{1}{3}$ and $\frac{1}{3}$, for $E$ and $E^c$ are respectively $-1$ and 1; and for $N$, $N^c$, and
$S$ are zero. The complete gauge symmetry breaking chains are 
\begin{align} 
                     & \SU(3)_C \times \SU(3)_L \times \SU(3)_R \nonumber\\ 
\xrightarrow{\text{Splitting}}\quad & \SU(3)_C \times \SU(2)_L \times \U(1)_{Y_{L}} \times \U(1)_{I_{3R}} \times \U(1)_{Y_{R}} \nonumber\\ 
\xrightarrow{\text{VEVs}}\quad      & \SU(3)_C \times \SU(2)_L \times \U(1)_Y\,. \label{symmetrybreaking}
\end{align}

The process of dynamical supersymmetry breaking has been studied in \cite{Cvetic:2003yd} for D6-brane models from Type IIA orientifolds.
The complex structure moduli $U$ can be obtained from the supersymmetry conditions as \cite{Sabir:2022hko},
\begin{align}\label{U-moduli}
U^i & = \frac{4i \chi^i+2\beta_i^2\chi_i^2}{4+\beta_i^2\chi_i^2}, \qquad \because \chi^i \equiv \frac{R_2^i}{R_1^i}.
\end{align}
These upper case moduli in string theory basis can be transformed in to lower case {$s$, $t$, $u$} moduli in field theory basis as \cite{Lust:2004cx},
\begin{equation}\label{eq:sugra-string-basis}
    \begin{split}
    \RealPart{s}   &= \frac{e^{-\phi_4}}{2\pi}\frac{\sqrt{\ImaginaryPart{U^1} \,  \ImaginaryPart{U^2}  \, \ImaginaryPart{U^3}  }   }{\Abs{U^1 U^2 U^3}  }\,, \\
    \RealPart{u^1} &= \frac{e^{-\phi_4}}{2\pi} \sqrt{\frac{\ImaginaryPart{U^1}  }{\ImaginaryPart{U^2}\,  \ImaginaryPart{U^3}}  } \Abs{\frac{U^2 U^3}{U^1}}\,, \\
    \RealPart{u^2} &= \frac{e^{-\phi_4}}{2\pi} \sqrt{\frac{\ImaginaryPart{U^2}  }{\ImaginaryPart{U^1}\,  \ImaginaryPart{U^3}}  } \Abs{\frac{U^1 U^3}{U^2}}\,, \\
    \RealPart{u^3} &= \frac{e^{-\phi_4}}{2\pi} \sqrt{\frac{\ImaginaryPart{U^3}  }{\ImaginaryPart{U^1}\,  \ImaginaryPart{U^2}}  } \Abs{\frac{U^1 U^2}{U^3}}\,.
    \end{split}
\end{equation}
where $\phi_4$ is the four dimensional dilaton which is related to the supergravity moduli as
\begin{equation}
2\pi e^{\phi_4}=\Big(\mathrm{Re}(s)\,\mathrm{Re}(u_1)\,\mathrm{Re}(u_2)\,\mathrm{Re}(u_3)\Big)^{-1/4}.
\end{equation}
Note that the three moduli parameter $\chi_1$, $\chi_2$, $\chi_3$ are not independent, as they can be expressed in terms of $x_A, x_B, x_C, x_D$ and the latter parameters are related by the supersymmetric condition~\eqref{susyconditions}. Actually, one can determine $\chi_1,\, \chi_2,\, \chi_3$ up to an overall coefficient, namely an action of dilation on these parameters. So one has to stabilize this dilation to determine all the moduli parameters. The holomorphic gauge kinetic function for any D6-brane stack $x$ wrapping a calibrated 3-cycle is given as \cite{Blumenhagen:2006ci},
\begin{equation}
f_x = \frac{1}{2\pi \ell_s^3}\left[e^{-\phi}\int_{\Pi_x} \mbox{Re}(e^{-i\theta_x}\Omega_3)-i\int_{\Pi_x}C_3\right],
\end{equation}
where the integral involving 3-form $\Omega_3$ gives,
\begin{equation}
\int_{\Pi_x}\Omega_3 = \frac{1}{4}\prod_{i=1}^3(n_x^iR_1^i + 2^{-\beta_i}il_x^iR_2^i).
\end{equation}
It can then be shown that,
\begin{align}
f_x &=
\frac{1}{4\kappa_x}(n_x^1\,n_x^2\,n_x^3\,s-\frac{n_x^1\,l_x^2\,l_x^3\,u^1}{2^{(\beta_2+\beta_3)}}-\frac{l_x^1\,n_x^2\,l_x^3\,u^2}{2^{(\beta_1+\beta_3)}}-
\frac{l_x^1\,l_x^2\,n_x^3\,u^3}{2^{(\beta_1+\beta_2)}}),
\label{kingauagefun}
\end{align}
where the factor ${\kappa}_x$ is the Kac-Moody level of the corresponding gauge Kac-Moody algebra associated to the D6-brane stacks such that ${\kappa}_x =1$ for $\U(N_x)$ and ${\kappa}_x =2$ for USp($2N_x$) or SO($2N_x$) \cite{Ginsparg:1987ee, Hamada:2014eia}.
The gauge coupling constant related to any stack $x$ of D6-branes is
\begin{equation}
    g_x^{-2} = \Abs{\RealPart{f_x}}\,,
\end{equation}
Since, the standard model hypercharge $\U(1)_Y$ is a linear combination of several $\U(1)$s as defined in \eqref{eq:QY}, the holomorphic gauge kinetic function for the hypercharge is also taken as a linear combination of the kinetic gauge functions from all of the stacks as \cite{Blumenhagen:2003jy, Ibanez:2001nd},
\begin{equation}\label{eq:fY}
f_Y= f_{b}+f_{c},
\end{equation}
where we have taken into account that our U(1)s are not canonically normalized. And the coupling constant for the hypercharge $g_Y$ can be determined as,
\begin{equation}
g^{-2}_Y = \Abs{\RealPart{f_Y}}\,.
\end{equation}
Hence, the tree-level gauge couplings have the relation,
\begin{equation}
g_a^2 = \alpha g_b^2 = \beta \frac{5}{3}g_Y^2 = \gamma (\pi e^{-\phi_4})
\end{equation}
where $\alpha$, $\beta$ and $\gamma$ denote the strengths of the strong coupling, the weak coupling, and the hypercharge coupling respectively.
Moreover, the K\"ahler potential takes the following form,
\begin{equation}\label{eq: Kaehler_potential}
    K = - \ln(S + \overline{S}) - \sum_{i=1}^3\ln(U^i + \overline{U}^i).
\end{equation}
And the one-loop beta functions for each hidden sector gauge group,
\begin{equation}\label{eq: beta_functions}
\begin{split}
    \beta^g_i &= - 3(\frac{N^{(i)}}{2} + 1) + 2 \Abs{I_{ai}} + \Abs{I_{bi}} + \Abs{I_{ci}} + 3(\frac{N^{(i)}}{2} - 1) \\
    & = -6 + 2\Abs{I_{ai} }+\Abs{I_{bi} } + \Abs{I_{ci}}  
\end{split}
\end{equation}
are required to be negative \cite{Cvetic:2004ui}.

Since we have large RR charges from three $\SU(3)$ groups, the construction becomes very difficult. Furthermore, unlike the Pati-Salam gauge symmetry where we avoid the nonvanishing torsion charges by taking an even number of D-branes, here in trinification gauge group, due to the appearance of three odd gauge factors of $\SU(3)$ the K-theory conditions \label{K-charges} turn out to be highly restrictive. Consequently, we were not able to construct any three-family trinification model for the $\T^6/(\Z_2\times \Z_2)$ compactification which requires atleast one two-torus to be tilted. 
We next turn our attention to the construction of four-family models which only require rectangular two-tori where RR tadpoles can be satisfied with lesser difficulty.

\section{Particle spectra of four-family supersymmetric trinification models}\label{sec:Pheno} 
Employing novel random and supervised scanning algorithm \cite{Li:2019nvi, Li:2021pxo} we have systematically scanned for the ${\cal N}=1$ supersymmetric trinification models with the gauge group $\SU(3)_C \times \SU(3)_L \times \SU(3)_R $. We have obtained some new models where the largest observed wrapping number is 2. After imposing various T-duality equivalence conditions \cite{Cvetic:2004ui}, there only remain three inequivalent models as listed in the appendix~\ref{Appendix}.  
 
\begin{table}[th]
\centering \footnotesize\renewcommand{\arraystretch}{1.3}
\caption{The spectrum of chiral and vector-like superfields, and their quantum numbers under the gauge symmetry $\SU(3)_C\times \SU(3)_L \times \SU(3)_R \times \U(1) \times \USp(8)$ for the Model~\hyperref[model1]{1}.\\}\label{tab:spec1}
$\begin{array}{|c||c||r|r|r||c|c|c|}\hline
\text{Model~\hyperref[model1]{1}} & \text{Quantum Number}  & Q_C & Q_{L} & Q_{R} & Q_{em} & B-L & \text{Field} \\
\hline\hline
ab               & 1 \times (3,\overline{3},1,1,1)           &  1 & -1 &  0  & -\frac{1}{3}, \frac{2}{3}, -1, 0 & \frac{1}{3}, -1  &  Q_L, L_L\\
ab'              & 3 \times (3,3,1,1,1)           &  1 & 1 &  0  & -\frac{1}{3}, \frac{2}{3}, -1, 0 & \frac{1}{3}, -1  &  Q_L, L_L\\
ac               & 1\times (\overline{3},1,3,1,1)            & -1 &  0 &  1  & \frac{1}{3}, -\frac{2}{3}, 1, 0  & -\frac{1}{3}, 1  &  Q_R, L_R\\
ac'              & 3\times (\overline{3},1,\overline{3},1,1)            & -1 &  0 &  -1  & \frac{1}{3}, -\frac{2}{3}, 1, 0  & -\frac{1}{3}, 1  &  Q_R, L_R\\
bc               & 8 \times (1,\overline{3},3,1,1)           &  0 & -1 &  1  & 1, 0, 0, -1 &  0  &  H\\
bd               & 3\times (1,\overline{3},1,1,1)            &  0 &  -1 &  0  & \mp \frac{1}{2} & 0  &    \\
bd'              & 3\times (1,\overline{3},1,\overline{1},1) &  0 & -1 &  0  & \mp \frac{1}{2} & 0  &    \\
cd               & 3\times (1,1,3,\overline{1},1)            &  0 &  0 &  1  & \pm \frac{1}{2} & 0  &    \\
cd'              & 3\times (1,1,3,1,1)&  0 &  0 & 1  & \pm \frac{1}{2} & 0  &    \\
b3               & 1\times (1,3,1,1,8)            &  0 &  1 &  0  & \pm \frac{1}{2} & 0  &    \\
c3               & 1\times (1,1,\overline{3},1,8)            &  0 &  1 &  0  & \mp \frac{1}{2} & 0  &    \\ 
b_{\yng(2)}              & 2\times(1,6,1,1,1)                &  0 &  2 &  0  & 0, \pm 1  &  0 &    \\
b_{\overline{\yng(1,1)}} & 2\times(1,3,1,1,1)     &  0 &  -2 & 0  & 0         &  0 &    \\
c_{\overline{\yng(2)}}              & 2\times(1,1,\overline{6},1,1)                &  0 &  0 &  -2  & 0, \pm 1  &  0 &    \\
c_{\yng(1,1)_{}} & 2\times(1,1,3,1,1)     &  0 &  0 & 2  & 0         &  0 &    \\ 
\hline
\end{array}$
\end{table} 

The detailed particle spectrum of Model~\hyperref[model1]{1} is shown in table~\ref{tab:spec1}. The model is self-dual under the exchange of $b$ and $c$ sectors.

\begin{table}[th]
\centering \footnotesize\renewcommand{\arraystretch}{1.3}
\caption{The spectrum of chiral and vector-like superfields, and their quantum numbers under the gauge symmetry $\SU(3)_C\times \SU(3)_L \times \SU(3)_R \times \U(1) \times \USp(2)^3$ for the Model~\hyperref[model2]{2}.\\}\label{tab:spec2}
$\begin{array}{|c||c||r|r|r||c|c|c|}\hline
\text{Model~\hyperref[model2]{2}} & \text{Quantum Number}  & Q_C & Q_{L} & Q_{R} & Q_{em} & B-L & \text{Field} \\
\hline\hline
ab               & 1 \times (3,\overline{3},1,1,1,1,1)           &  1 & -1 &  0  & -\frac{1}{3}, \frac{2}{3}, -1, 0 & \frac{1}{3}, -1  &  Q_L, L_L\\
ab'              & 3 \times (3,3,1,1,1,1,1)           &  1 & 1 &  0  & -\frac{1}{3}, \frac{2}{3}, -1, 0 & \frac{1}{3}, -1  &  Q_L, L_L\\
ac               & 1\times (\overline{3},1,3,1,1,1,1)            & -1 &  0 &  1  & \frac{1}{3}, -\frac{2}{3}, 1, 0  & -\frac{1}{3}, 1  &  Q_R, L_R\\
ac'              & 3\times (\overline{3},1,\overline{3},1,1,1,1)            & -1 &  0 &  -1  & \frac{1}{3}, -\frac{2}{3}, 1, 0  & -\frac{1}{3}, 1  &  Q_R, L_R\\
bc               & 4 \times (1,\overline{3},3,1,1,1,1)           &  0 & -1 &  1  & 1, 0, 0, -1 &  0  &  H\\
bd               & 3\times (1,3,1,\overline{1},1,1,1)            &  0 &  1 &  0  & \pm \frac{1}{2} & 0  &    \\
bd'              & 3\times (1,3,1,1,1,1,1) &  0 & 1 &  0  & \pm \frac{1}{2} & 0  &    \\
cd               & 18\times (1,1,3,\overline{1},1,1,1)            &  0 &  0 &  1  & \pm \frac{1}{2} & 0  &    \\
cd'              & 10\times (1,1,\overline{3},\overline{1},1,1,1)&  0 &  0 & -1  & \mp \frac{1}{2} & 0  &    \\
ad               & 5\times (3,1,1,\overline{1},1,1,1)            &  1 &  0 &  0  &  \frac{1}{6}, -\frac{1}{2} & \frac{1}{3}, -1  &    \\
ad'              & 9\times (3,1,1,1,1,1,1)&  1 &  0 & 0  & \frac{1}{6}, -\frac{1}{2} & \frac{1}{3}, -1  &    \\
a1               & 1\times (\overline{3},1,1,1,2,1,1)            &  -1 &  0 &  0  & -\frac{1}{6}, \frac{1}{2} & -\frac{1}{3},1  &    \\
b1               & 1\times (1,\overline{3},1,1,2,1,1)            &  0 &  -1 &  0  & \mp \frac{1}{2} & 0  &    \\
d1               & 8\times (1,1,1,\overline{1},2,1,1)            &  0 &  0 &  0  & 0 & 0  &    \\
a3               & 2\times (3,1,1,1,1,2,1)            &  1 &  0 &  0  & \frac{1}{6},-\frac{1}{2} & \frac{1}{3},-1  &    \\
c3               & 2\times (1,1,3,1,1,2,1)            &  0 &  0 &  1  & \pm \frac{1}{2} & 0  &    \\ 
b4               & 1\times (1,3,1,1,1,1,2)            &  0 &  1 &  0  & \pm \frac{1}{2} & 0  &    \\
c4               & 1\times (1,1,\overline{3},1,1,1,2)            &  0 &  0 &  -1  & \mp \frac{1}{2} & 0  &    \\
d4               & 2\times (1,1,1,\overline{1},1,1,2)            &  0 &  0 &  0  & 0 & 0  &    \\
a_{\overline{\yng(2)}}              & 2\times(\overline{6},1,1,1,1,1,1)                &  -2 &  0 &  0  & -\frac{1}{3}, \frac{1}{3},1  &  -\frac{2}{3},2 &    \\
a_{\yng(1,1)} & 2\times(3,1,1,1,1,1,1)     &  2 &  0 & 0  & \frac{1}{3}, -1  &  \frac{2}{3},-2 &    \\
c_{\overline{\yng(2)}}              & 2\times(1,1,\overline{6},1,1,1,1)                &  0 &  0 &  -2  & 0, \pm 1  &  0 &    \\
c_{\yng(1,1)_{}} & 2\times(1,1,3,1,1,1,1)     &  0 &  0 & 2  & 0         &  0 &    \\ 
\hline
\end{array}$
\end{table} 
 
The complete particle spectra of models~\hyperref[model2]{2} and \hyperref[model2dual]{2-dual} are listed in tables~\ref{tab:spec2} and \ref{tab:spec2dual} respectively. The models are dual under the exchange of $b$ and $c$ sectors.  

\begin{table}[th]
\centering \footnotesize\renewcommand{\arraystretch}{1.3}
\caption{The spectrum of chiral and vector-like superfields, and their quantum numbers under the gauge symmetry $\SU(3)_C\times \SU(3)_L \times \SU(3)_R \times \U(1) \times \USp(2)^3$ for the Model~\hyperref[model2dual]{2-dual}.\\}\label{tab:spec2dual}
$\begin{array}{|c||c||r|r|r||c|c|c|}\hline
\text{Model~\hyperref[model2dual]{2-dual}} & \text{Quantum Number}  & Q_C & Q_{L} & Q_{R} & Q_{em} & B-L & \text{Field} \\
\hline\hline
ab               & 1 \times (\overline{3},3,1,1,1,1,1)           &  -1 & 1 &  0  & \frac{1}{3}, -\frac{2}{3}, 1, 0  & -\frac{1}{3}, 1  &  Q_L, L_L\\
ab'              & 3 \times (\overline{3},\overline{3},1,1,1,1,1)           &  -1 & -1 &  0  & \frac{1}{3}, -\frac{2}{3}, 1, 0 & -\frac{1}{3}, 1  &  Q_L, L_L\\
ac               & 1\times (3,1,\overline{3},1,1,1,1)            & -1 &  0 &  1  & -\frac{1}{3}, \frac{2}{3}, -1, 0 & \frac{1}{3}, -1  &  Q_R, L_R\\
ac'              & 3\times (3,1,3,1,1,1,1)            & 1 &  0 &  1  & -\frac{1}{3}, \frac{2}{3}, -1, 0  & \frac{1}{3}, -1  &  Q_R, L_R\\
bc               & 4 \times (1,3,\overline{3},1,1,1,1)           &  0 & 1 &  -1  & 1, 0, 0, -1 &  0  &   H\\
cd               & 3\times (1,1,3,\overline{1},1,1,1)            &  0 &  0 &  1  & \pm \frac{1}{2} & 0  &    \\
cd'              & 3\times (1,1,3,1,1,1,1) &  0 & 0 &  1  & \pm \frac{1}{2} & 0  &    \\
bd               & 18\times (1,3,1,\overline{1},1,1,1)            &  0 & 1 &  0  & \pm \frac{1}{2} & 0  &    \\
bd'              & 10\times (1,\overline{3},1,\overline{1},1,1,1)&  0 &  -1 & 0  & \mp \frac{1}{2} & 0  &    \\
ad               & 5\times (3,1,1,\overline{1},1,1,1)            &  1 &  0 &  0  &  \frac{1}{6}, -\frac{1}{2} & \frac{1}{3}, -1  &    \\
ad'              & 9\times (3,1,1,1,1,1,1)&  1 &  0 & 0  & \frac{1}{6}, -\frac{1}{2} & \frac{1}{3}, -1  &    \\
a1               & 1\times (\overline{3},1,1,1,2,1,1)            &  -1 &  0 &  0  & -\frac{1}{6}, \frac{1}{2} & -\frac{1}{3},1  &    \\
c1               & 1\times (1,1,\overline{3},1,2,1,1)            &  0 &  0 &  -1  & \mp \frac{1}{2} & 0  &    \\
d1               & 8\times (1,1,1,\overline{1},2,1,1)            &  0 &  0 &  0  & 0 & 0  &    \\
a3               & 2\times (3,1,1,1,1,2,1)            &  1 &  0 &  0  & \frac{1}{6},-\frac{1}{2} & \frac{1}{3},-1  &    \\
b3               & 2\times (1,3,1,1,1,2,1)            &  0 &  1 &  0  & \pm \frac{1}{2} & 0  &    \\ 
c4               & 1\times (1,1,3,1,1,1,2)            &  0 &  0 &  1  & \pm \frac{1}{2} & 0  &    \\
b4               & 1\times (1,\overline{3},1,1,1,1,2)            &  0 &  -1 &  0  & \mp \frac{1}{2} & 0  &    \\
d4               & 2\times (1,1,1,\overline{1},1,1,2)            &  0 &  0 &  0  & 0 & 0  &    \\
a_{\overline{\yng(2)}}              & 2\times(\overline{6},1,1,1,1,1,1)                &  -2 &  0 &  0  & -\frac{1}{3}, \frac{1}{3},1  &  -\frac{2}{3},2 &    \\
a_{\yng(1,1)} & 2\times(3,1,1,1,1,1,1)     &  2 &  0 & 0  & \frac{1}{3}, -1  &  \frac{2}{3},-2 &    \\
b_{\overline{\yng(2)}}              & 2\times(1,\overline{6},1,1,1,1,1)                &  0 &  -2 &  0  & 0, \pm 1  &  0 &    \\
b_{\yng(1,1)_{}} & 2\times(1,3,1,1,1,1,1)     &  0 &  2 & 0  & 0         &  0 &    \\ 
\hline
\end{array}$
\end{table}

\begin{table}[th]
\centering \footnotesize\renewcommand{\arraystretch}{1.3}
\caption{The spectrum of chiral and vector-like superfields, and their quantum numbers under the gauge symmetry $\SU(3)_C\times \SU(3)_L \times \SU(3)_R \times \USp(2) \times \USp(6)$ for the Model~\hyperref[model3]{3}.\\}\label{tab:spec3}
$\begin{array}{|c||c||r|r|r||c|c|c|}\hline
\text{Model~\hyperref[model3]{3}} & \text{Quantum Number}  & Q_C & Q_{L} & Q_{R} & Q_{em} & B-L & \text{Field} \\
\hline\hline
ab               & 3 \times (\overline{3},3,1,1,1,1)           &  -1 & 1 &  0  & -\frac{1}{3}, \frac{2}{3}, -1, 0  & \frac{1}{3}, -1  &  Q_L, L_L\\
ab'              & 1 \times (\overline{3},\overline{3},1,1,1,1) &  -1 & -1 &  0  & -\frac{1}{3}, \frac{2}{3}, -1, 0 & \frac{1}{3}, -1  &  Q_L, L_L\\
ac               & 3\times (3,1,\overline{3},1,1,1)            & 1 &  0 &  -1  & \frac{1}{3}, -\frac{2}{3}, 1, 0 & -\frac{1}{3}, 1  &  Q_R, L_R\\
ac'              & 1\times (3,1,3,1,1,1)                       & 1 &  0 &  1  & \frac{1}{3}, -\frac{2}{3}, 1, 0  & -\frac{1}{3}, 1  &  Q_R, L_R\\
bc               & 4 \times (1,\overline{3},3,1,1,1)           &  0 & -1 &  1  & 1, 0, 0, -1 &  0  &  H\\
ad               & 3\times (\overline{3},1,1,2,1,1)            &  -1 &  0 &  0  &  \frac{1}{6}, -\frac{1}{2} & \frac{1}{3}, -1  &    \\
ad'              & 3\times (\overline{3},1,1,\overline{2},1,1)  &  -1 &  0 & 0  & \frac{1}{6}, -\frac{1}{2} & \frac{1}{3}, -1  &    \\
cd               & 6\times (1,1,3,\overline{1},1,1)            &  0 &  0 &  1  & \pm \frac{1}{2} & 0  &    \\
a2               & 2\times (\overline{3},1,1,1,2,1)            & -1 &  0 &  0  & \frac{1}{6},-\frac{1}{2} & \frac{1}{3},-1  &    \\
b3               & 1\times (1,\overline{3},1,1,1,6)            &  0 &  -1 &  0  & \pm \frac{1}{2} & 0  &    \\
c2               & 2\times (1,1,2,1,\overline{2},1)            &  0 &  0 &  1  & \pm \frac{1}{2} & 0  &    \\
c3               & 1\times (1,1,\overline{2},1,1,6)            &  0 &  0 &  -1  & \mp \frac{1}{2} & 0  &    \\
d2               & 1\times (1,1,1,\overline{2},2,1)            &  0 &  0 &  0  & 0 & 0  &    \\
d3               & 2\times (1,1,1,\overline{2},1,6)            &  0 &  0 &  0  & 0 & 0  &    \\ 
a_{\yng(2)}      & 2\times(6,1,1,1,1,1,1)                &  2 &  0 &  0  & -\frac{1}{3}, \frac{1}{3},1  &  -\frac{2}{3},2 &    \\
a_{\overline{\yng(1,1)}} & 2\times(3,1,1,1,1,1,1)     &  -2 &  0 & 0  & \frac{1}{3}, -1  &  \frac{2}{3},-2 &    \\
c_{\overline{\yng(2)}}              & 2\times(1,1,\overline{6},1,1,1)                &  0 &  0 &  -2  & 0, \pm 1  &  0 &    \\
c_{\yng(1,1)} & 2\times(1,1,3,1,1,1)     &  0 &  0 & 2  & 0         &  0 &    \\ 
d_{\overline{\yng(1,1)}_{}} & 16\times(1,1,1,\overline{1},1,1) &  0 &  0 &  0  & 0         &  0 &    \\
\hline
\end{array}$
\end{table}

\begin{table}[th]
\centering \footnotesize\renewcommand{\arraystretch}{1.3}
\caption{The spectrum of chiral and vector-like superfields, and their quantum numbers under the gauge symmetry $\SU(3)_C\times \SU(3)_L \times \SU(3)_R \times \USp(2) \times \USp(6)$ for the Model~\hyperref[model3dual]{3-dual}.\\}\label{tab:spec3dual}
$\begin{array}{|c||c||r|r|r||c|c|c|}\hline
\text{Model~\hyperref[model3dual]{3-dual}} & \text{Quantum Number}  & Q_C & Q_{L} & Q_{R} & Q_{em} & B-L & \text{Field} \\
\hline\hline
ab               & 3 \times (3,\overline{3},1,1,1,1)           &  1 & -1 &  0  & \frac{1}{3}, -\frac{2}{3}, 1, 0  & -\frac{1}{3}, 1  &  Q_L, L_L\\
ab'              & 1 \times (3,3,1,1,1,1) &  1 & 1 &  0  & -\frac{1}{3}, \frac{2}{3}, -1, 0 & \frac{1}{3}, -1  &  Q_L, L_L\\
ac               & 3\times (\overline{3},1,3,1,1,1)            & -1 &  0 &  1  & -\frac{1}{3}, \frac{2}{3}, -1, 0 & \frac{1}{3}, -1  &  Q_R, L_R\\
ac'              & 1\times (\overline{3},1,\overline{3},1,1,1)                       & 1 &  0 &  1  & \frac{1}{3}, -\frac{2}{3}, 1, 0  & -\frac{1}{3}, 1  &  Q_R, L_R\\
bc               & 4 \times (1,3,\overline{3},1,1,1)           &  0 & 1 &  -1  & 1, 0, 0, -1 &  0  &  H\\
ad               & 3\times (\overline{3},1,1,2,1,1)            &  -1 &  0 &  0  &  \frac{1}{6}, -\frac{1}{2} & \frac{1}{3}, -1  &    \\
ad'              & 3\times (\overline{3},1,1,\overline{2},1,1)  &  -1 &  0 & 0  & \frac{1}{6}, -\frac{1}{2} & \frac{1}{3}, -1  &    \\
bd               & 6\times (1,3,1,\overline{1},1,1)            &  0 &  0 &  1  & \pm \frac{1}{2} & 0  &    \\
a2               & 2\times (\overline{3},1,1,1,2,1)            & -1 &  0 &  0  & \frac{1}{6},-\frac{1}{2} & \frac{1}{3},-1  &    \\
c3               & 1\times (1,1,\overline{3},1,1,6)            &  0 &  -1 &  0  & \pm \frac{1}{2} & 0  &    \\
b2               & 2\times (1,2,1,1,\overline{2},1)            &  0 &  0 &  1  & \pm \frac{1}{2} & 0  &    \\
b3               & 1\times (1,\overline{2},1,1,1,6)            &  0 &  0 &  -1  & \mp \frac{1}{2} & 0  &    \\
d2               & 1\times (1,1,1,\overline{2},2,1)            &  0 &  0 &  0  & 0 & 0  &    \\
d3               & 2\times (1,1,1,\overline{2},1,6)            &  0 &  0 &  0  & 0 & 0  &    \\ 
a_{\yng(2)}      & 2\times(6,1,1,1,1,1,1)                &  2 &  0 &  0  & -\frac{1}{3}, \frac{1}{3},1  &  -\frac{2}{3},2 &    \\
a_{\overline{\yng(1,1)}} & 2\times(3,1,1,1,1,1,1)     &  -2 &  0 & 0  & \frac{1}{3}, -1  &  \frac{2}{3},-2 &    \\
b_{\overline{\yng(2)}}              & 2\times(1,\overline{6},1,1,1,1)                &  0 &  0 &  -2  & 0, \pm 1  &  0 &    \\
b_{\yng(1,1)} & 2\times(1,3,1,1,1,1)     &  0 &  0 & 2  & 0         &  0 &    \\ 
d_{\overline{\yng(1,1)}_{}} & 16\times(1,1,1,\overline{1},1,1) &  0 &  0 &  0  & 0         &  0 &    \\
\hline
\end{array}$
\end{table}  

The complete particle spectra of models~\hyperref[model3]{3} and \hyperref[model3dual]{3-dual} are listed in tables~\ref{tab:spec3} and \ref{tab:spec3dual} respectively. The models are dual under the exchange of $b$ and $c$ sectors.

All models feature several exotic states that need to be decoupled at some higher energy scale for the models to be realistic.

\FloatBarrier

\section{Conclusion}\label{sec:conclusion}
We have revisited the construction of $\mathcal{N}=1$ supersymmetric trinification grand unified models arising from intersecting D6-branes at angles on a $\T^6/(\Z_2\times \Z_2)$ orientifold from IIA string theory. We have discussed the gauge symmetry breaking from the trinification symmetry down to the standard model gauge group. Notably, due to large RR charges from product of three $\SU(3)$ gauge factors and the associated non-trivial K-theory constraints we did not find any three-family trinification models in our search. However, we have found three inequivalent four-family models where the highest wrapping number observed is 2. We have also listed the detailed perturbative particle spectra and the tree-level gauge coupling relations at the string-scale. All models feature exotic states from the hidden sector.


\appendix

\section{Four-family supersymmetric trinification models}\label{Appendix}

In the appendix, we list all representative four-family supersymmetric trinification models obtained from random scanning method. $a, b, c, d$ in the first column in every table represent the four stacks of D6-branes, respectively. Similarly, $1, 2, 3, 4$ in the first columns is a short-handed notation for the filler branes along the $\Omega {\cal R}$, $\Omega {\cal R} \omega$, $\Omega {\cal R} \theta \omega$ and $\Omega {\cal R} \theta$ O6-planes, respectively. The second column in each table lists the numbers of D6-branes in the respective stack. In the third column we record the wrapping numbers of each D6-brane configuration. The rest of the columns record the intersection numbers between various stacks. For instance, in the $b$ column of table~\ref{model1}, from top to bottom, the numbers represent intersection numbers $I_{ab}, I_{bc}, I_{bd}$, \textit{etc.}.  As usual, $b'$ and $c'$ are the orientifold $\Omega {\cal R}$ image of $b$ and $c$ stacks of D6-branes. We also list the relation between $x_A, x_B, x_C, x_D$, which are determined by the supersymmetry conditions~\eqref{susyconditions}, as well as the relation between the moduli parameter $\chi_1,\, \chi_2,\, \chi_3$. The one loop $\beta$ functions $\beta^g_i$ for each filler brane stack is also listed. The gauge coupling relations are listed in the caption of each table.

\begin{table}[th]  

\centering \footnotesize
\caption{D6-brane configurations and intersection numbers of Model~\hyperref[model1]{1}, and its gauge coupling relation is $g_a^2=\frac{14 g_b^2}{5}=\frac{14 g_c^2}{5}=\frac{84}{25} \frac{5 g_Y^2}{3}=\frac{8}{5} 6^{3/4} \pi  e^{\text{$\phi $4}}$} \label{model1}
$\begin{array}{|c|c|c| r|r|r| r|r|r| r|r|r| r|r|r|r|}
	\hline
	\multicolumn{2}{|c|}{\text{Model~\hyperref[model1]{1}}}  & \multicolumn{13}{c|}{\U(3)_C\times \U(3)_L \times \U(3)_R \times \U(1) \times \USp(8)} \\
	\hline \hline \rm{stack} & N & (n^1,l^1)\times (n^2,l^2)\times (n^3,l^3) & n_{\yng(2)} & n_{\yng(1,1)_{}} & b & b' & c & c'& d & d' & 1 & 2 & 3 & 4 \\
	\hline
	a   & 6   & (0, -1)\times (-1, -1)\times (-1, -1) & 0    & 0     & 1    & 3  & -1  & -3     & 0     & 0     & 0     & 0     & 0    & 0 \\
    b   & 6   & (-1, 1)\times (2, 1)\times (-1, 0)      & 2     & -2     & \text{-} &  \text{-} & -8     & 0    & -3     & -3    & 0     & 0     & 1    & 0 \\
    c   & 6   & (-1, -1)\times (0, 1)\times (1, 2)    & -2     & 2    & \text{-} & \text{-}  & \text{-} & \text{-}  & 3     & 3   & 0     & 0     & -1    & 0 \\
	\hline 
     d   & 2   & (-2, -1)\times (-1, -1)\times (1, -1) &  \multicolumn{12}{c|}{x_A=x_B=\frac{1}{2}x_C=\frac{1}{3}x_D} \\
     3   &   8  & (0, -1)\times (1, 0)\times (0, 1) &  \multicolumn{12}{c|}{\chi_1=1/\sqrt{6}, \quad \chi_2=\sqrt{6}/3, \quad \chi_3= \sqrt{6}/2,} \\
         &     &                     &  \multicolumn{12}{c|}{\beta^g_d=0, \quad\beta^g_3=-4} \\
	\hline
\end{array}$
\end{table}


\begin{table}[th]  
\centering \footnotesize
\caption{D6-brane configurations and intersection numbers of Model~\hyperref[model2]{2}, and its gauge coupling relation is $g_a^2=\frac{8 g_b^2}{5}=\frac{8 g_c^2}{5}=\frac{48}{25} \frac{5 g_Y^2}{3}=\frac{8}{15} \sqrt{2} 11^{3/4} \pi  e^{\text{$\phi $4}}$.\\} \label{model2}
$\begin{array}{|c|c|c| r|r|r| r|r|r| r|r|r| r|r|r|r|}
	\hline
	\multicolumn{2}{|c|}{\text{Model~\hyperref[model2]{2}}}  & \multicolumn{13}{c|}{\U(3)_C\times \U(3)_L \times \U(3)_R \times \U(1)\times \USp(2)^3} \\
	\hline \hline \rm{stack} & N & (n^1,l^1)\times (n^2,l^2)\times (n^3,l^3) & n_{\yng(2)} & n_{\yng(1,1)_{}} & b & b' & c & c'& d & d' & 1 & 2 & 3 & 4 \\
	\hline
	a   & 6   & (-2, 1)\times (0, 1)\times (-1, 1) & -2    & 2     & 1    & 3  & -1  & -3     & 5     & 9     & -1     & 0     & 2    & 0 \\
    b   & 6   & (1, -1)\times (1, -1)\times (0, 1)      & 0     & 0     & \text{-} &  \text{-} & -4     & 0    & 3     & 3    & -1     & 0     & 0    & 1 \\
    c   & 6   & (-1, 0)\times (1, 1)\times (-2, 1)    & -2     & 2    & \text{-} & \text{-}  & \text{-} & \text{-}  & 18     & -10   & 0     & 0     & 2    & -1 \\
    \hline
    d   & 2   & (1, 2)\times (1, -2)\times (1, -2)  & \multicolumn{12}{c|}{x_A=\frac{2}{11}x_B=2x_C=x_D} \\
       1   &   2  & (1, 0)\times (1, 0)\times (1, 0) &  \multicolumn{12}{c|}{\chi_1=\sqrt{11}, \quad \chi_2=1/\sqrt{11}, \quad \chi_3= 2/\sqrt{11}} \\
          3   &   2  & (0, -1)\times (1, 0)\times (0, 1) &  \multicolumn{12}{c|}{ \beta^g_d=25, \quad \beta^g_1=-3, \quad \beta^g_3=0, \quad \beta^g_4=-4}\\
          4   &   2  & (0, -1)\times (0, 1)\times (1, 0) & \multicolumn{12}{c|}{ }\\
	\hline
\end{array}$
\end{table}


\begin{table}[th]  
\centering \footnotesize
\caption{D6-brane configurations and intersection numbers of Model~\hyperref[model2dual]{2-dual}, and its gauge coupling relation is $g_a^2=\frac{8 g_b^2}{5}=\frac{8 g_c^2}{5}=\frac{48}{25} \frac{5 g_Y^2}{3}=\frac{8}{15} \sqrt{2} 11^{3/4} \pi  e^{\text{$\phi $4}}$.\\} \label{model2dual}
$\begin{array}{|c|c|c| r|r|r| r|r|r| r|r|r| r|r|r|r|}
	\hline
	\multicolumn{2}{|c|}{\text{Model~\hyperref[model2dual]{2-dual}}}  & \multicolumn{13}{c|}{\U(3)_C\times \U(3)_L \times \U(3)_R \times \U(1)\times \USp(2)^3} \\
	\hline \hline \rm{stack} & N & (n^1,l^1)\times (n^2,l^2)\times (n^3,l^3) & n_{\yng(2)} & n_{\yng(1,1)_{}} & b & b' & c & c'& d & d' & 1 & 2 & 3 & 4 \\
	\hline
	a   & 6   & (-2, 1)\times (0, 1)\times (-1, 1) & -2    & 2     & -1    & -3  & 1  & 3     & 5     & 9     & -1     & 0     & 2    & 0 \\
    b   & 6   & (-1, 0)\times (1, 1)\times (-2, 1)    & -2     & 2    & \text{-} & \text{-}  & 4 & 0  & 18     & -10   & 0     & 0     & 2    & -1 \\
    c   & 6   & (1, -1)\times (1, -1)\times (0, 1)      & 0     & 0     & \text{-} &  \text{-} & \text{-}     & \text{-}    & 3     & 3    & -1     & 0     & 0    & 1 \\
    \hline
    d   & 2   & (1, 2)\times (1, -2)\times (1, -2)  & \multicolumn{12}{c|}{x_A=\frac{2}{11}x_B=2x_C=x_D} \\
       1   &   2  & (1, 0)\times (1, 0)\times (1, 0) &  \multicolumn{12}{c|}{\chi_1=\sqrt{11}, \quad \chi_2=1/\sqrt{11}, \quad \chi_3= 2/\sqrt{11}} \\
          3   &   2  & (0, -1)\times (1, 0)\times (0, 1) &  \multicolumn{12}{c|}{ \beta^g_d=25, \quad \beta^g_1=-3, \quad \beta^g_3=0, \quad \beta^g_4=-4}\\
          4   &   2  & (0, -1)\times (0, 1)\times (1, 0) & \multicolumn{12}{c|}{ }\\
	\hline
\end{array}$
\end{table}


\begin{table}[th]  
\centering \footnotesize
\caption{D6-brane configurations and intersection numbers of Model~\hyperref[model3]{3}, and its gauge coupling relation is $g_a^2=\frac{20 g_b^2}{13}=\frac{20 g_c^2}{13}=\frac{24}{13} \frac{5 g_Y^2}{3}=\frac{24}{13} \sqrt{6} \pi  e^{\text{$\phi $4}}$.\\} \label{model3}
$\begin{array}{|c|c|c| r|r|r| r|r|r| r|r|r| r|r|r|r|}
	\hline
	\multicolumn{2}{|c|}{\text{Model~\hyperref[model3]{3}}}  & \multicolumn{13}{c|}{\U(3)_C\times \U(3)_L \times \U(3)_R \times \U(2)\times \USp(2)\times \USp(6)} \\
	\hline \hline \rm{stack} & N & (n^1,l^1)\times (n^2,l^2)\times (n^3,l^3) & n_{\yng(2)} & n_{\yng(1,1)_{}} & b & b' & c & c'& d & d' & 1 & 2 & 3 & 4 \\
	\hline
	a   & 6   & (0, 1)\times (-1, -1)\times (2, 1)  & 2    & -2    & -3    & -1  & 3  & 1     & -3     & -3    &  0   &   -2    & 0    & 0 \\
    b   & 6   & (-1, 1)\times (0, 1)\times (-1, 1)  & 0    & 0     & \text{-} & \text{-}  & -4 & 0  & 0   & 0   & 0     & 0     & 1    & 0 \\
    c   & 6   & (-1, -1)\times (-2, 1)\times (1, 0) & -2   & 2     & \text{-} &  \text{-} & \text{-}     & \text{-}    & 6     & 0   & 0     & 2     & -1    & 0 \\
    d   & 4   & (-1, 1)\times (-1, 2)\times (1, 1)  & 0    & -16   & \text{-} & \text{-}  & \text{-}  & \text{-}  & \text{-} & \text{-}  & 0     & -1     & -2    & 0 \\
	\hline
       2   &   2  & (1, 0)\times (0, -1)\times (0, 1) &  \multicolumn{12}{c|}{x_A=2x_B=x_C=\frac{2}{9}x_D} \\
       3   &   6  & (0, -1)\times (1, 0)\times (0, 1) &  \multicolumn{12}{c|}{\chi_1=\frac{1}{3}, \quad \chi_2=\frac{2}{3}, \quad \chi_3= 3 } \\
           &      &                                   &  \multicolumn{12}{c|}{ \beta^g_d=6, \quad \beta^g_2=0, \quad \beta^g_3=-4}\\
	\hline
\end{array}$
\end{table}


\begin{table}[th]  
\centering \footnotesize
\caption{D6-brane configurations and intersection numbers of Model~\hyperref[model3dual]{3-dual}, and its gauge coupling relation is $g_a^2=\frac{20 g_b^2}{13}=\frac{20 g_c^2}{13}=\frac{24}{13} \frac{5 g_Y^2}{3}=\frac{24}{13} \sqrt{6} \pi  e^{\text{$\phi $4}}$.\\} \label{model3dual}
$\begin{array}{|c|c|c| r|r|r| r|r|r| r|r|r| r|r|r|r|}
	\hline
	\multicolumn{2}{|c|}{\text{Model~\hyperref[model3dual]{3-dual}}}  & \multicolumn{13}{c|}{\U(3)_C\times \U(3)_L \times \U(3)_R \times \U(2)\times \USp(2)\times \USp(6)} \\
	\hline \hline \rm{stack} & N & (n^1,l^1)\times (n^2,l^2)\times (n^3,l^3) & n_{\yng(2)} & n_{\yng(1,1)_{}} & b & b' & c & c'& d & d' & 1 & 2 & 3 & 4 \\
	\hline
	a   & 6   & (0, 1)\times (-1, -1)\times (2, 1)  & 2    & -2    & 3    & 1  & -3  & -1     & -3     & -3    &  0   &   -2    & 0    & 0 \\
    b   & 6   & (-1, -1)\times (-2, 1)\times (1, 0) & -2   & 2     & \text{-} &  \text{-} & 4     & 0    & 6     & 0   & 0     & 2     & -1    & 0 \\
    c   & 6   & (-1, 1)\times (0, 1)\times (-1, 1)  & 0    & 0     & \text{-} & \text{-}  & \text{-} & \text{-}  & 0   & 0   & 0     & 0     & 1    & 0 \\
    d   & 4   & (-1, 1)\times (-1, 2)\times (1, 1)  & 0    & -16   & \text{-} & \text{-}  & \text{-}  & \text{-}  & \text{-} & \text{-}  & 0     & -1     & -2    & 0 \\
	\hline
       2   &   2  & (1, 0)\times (0, -1)\times (0, 1) &  \multicolumn{12}{c|}{x_A=2x_B=x_C=\frac{2}{9}x_D} \\
       3   &   6  & (0, -1)\times (1, 0)\times (0, 1) &  \multicolumn{12}{c|}{\chi_1=\frac{1}{3}, \quad \chi_2=\frac{2}{3}, \quad \chi_3= 3 } \\
           &      &                                   &  \multicolumn{12}{c|}{ \beta^g_d=6, \quad \beta^g_2=0, \quad \beta^g_3=-4}\\
	\hline
\end{array}$
\end{table}

\FloatBarrier
\bibliographystyle{JHEP}

\providecommand{\href}[2]{#2}\begingroup\raggedright\endgroup

\end{document}